\documentclass[doublecol,figures]{epl2}

\usepackage{amsmath,amssymb}
\usepackage{hyperref,breakurl}
\usepackage{mciteplus}

\newcommand{\Uset}{\mathcal{U}}
\newcommand{\Oset}{\mathcal{O}}
\newcommand{\vek}[1]{\boldsymbol{#1}}

\title{
  The effect of discrete vs.\ continuous-valued ratings on reputation and
  ranking systems
}

\author{
  Mat{\'u}{\v s} Medo\inst{1}\thanks{E-mail: \email{matus.medo@unifr.ch}}
  \and
  Joseph Rushton Wakeling\inst{1,2,3}\thanks{E-mail: \email{joseph.wakeling@create-net.org}}
}
\shortauthor{Mat{\'u}{\v s} Medo and Joseph Rushton Wakeling}
\institute{
  \inst{1}D{\'e}partement de Physique, Universit{\'e} de Fribourg - Chemin du Mus{\'e}e 3, CH-1700~Fribourg, Switzerland \\
  \inst{2}Institut Jean Nicod (CNRS) - {\'E}cole Normale Sup{\'e}rieure, 29 rue d'Ulm, F-75005 Paris, France \\
  \inst{3}CREATE-NET Research Consortium - via alla Cascata 56D, 38123 Povo di Trento, Italy
}

\pacs{89.20.Ff}{Computer science and technology}
\pacs{89.65.-s}{Social and economic systems}
\pacs{89.20.Hh}{World Wide Web, Internet}

\abstract{
  When users rate objects, a sophisticated algorithm that takes into account
  ability or reputation may produce a fairer or more accurate aggregation of
  ratings than the straightforward arithmetic average.  Recently a number of
  authors have proposed different \emph{co-determination} algorithms where
  estimates of user and object reputation are refined iteratively together,
  permitting accurate measures of both to be derived directly from the rating
  data.  However, simulations demonstrating these methods' efficacy assumed a
  continuum of rating values, consistent with typical physical modelling
  practice, whereas in most actual rating systems only a limited range of
  discrete values (such as a 5-star system) is employed.
  We perform a comparative test of several co-determination algorithms with
  different scales of discrete ratings and show that this seemingly minor
  modification in fact has a significant impact on algorithms' performance.
  Paradoxically, where rating resolution is low, increased noise in users'
  ratings may even improve the overall performance of the system.
}

\date{8 March 2010}

\begin{document}

\maketitle

\section{Introduction}
With the growth of the internet and e-commerce~\cite{howmuchinfo}, an
increasing number of our social and commercial interactions are now one-shot
exchanges with strangers identifiable only by easily-replaced
pseudonyms~\cite{friedman2001}.  Similarly, most items on sale from e-commerce
websites must be purchased without an opportunity to try them first, creating
an information asymmetry that encourages the provision of low-quality
goods~\cite{melnik2002,*resnick2002,*resnick2006,*akerlof1970}.  To offset this
risk of fraud or deception, many online services implement \emph{reputation
systems}~\cite{resnick2000,*bolton2004,*masum2004,*josang2007} that collect
ratings and feedback from users so as to provide a measure of trustworthiness
for goods or individuals.

A key challenge is how to aggregate this feedback effectively given that not
all ratings are equal.  Some users' judgement may be poor or malicious: for
example, many eBay users forgo issuing deserved negative feedback to cheaters
because the negative feedback they will receive in reprisal will devastate
their own carefully cultivated good reputation~\cite{kollock1999}.  An
effective reputation system thus needs to distinguish between good and bad
raters and ratings.

One approach to this has been the development of \emph{co-determination}
algorithms of reputation, where aggregate reputation (or quality) of rated
objects\footnote{
  We use `object' simply as a generic term: the object of the rating.  This
  might be an actual object, such as a book or CD, or it might be a person or
  organization, such as an eBay auctioneer, a website, or an Amazon Marketplace
  seller.
} is used to estimate a corresponding reputation (or ability) for the system's
users, and this latter measure is then used to re-weight the aggregation of
ratings for objects~\cite{mizzaro2003,yu2006,*laureti2006,dekerchove2007,*dekerchove2008,*dekerchove2009,*dekerchove2010}.
By iterating this procedure over time, ratings from malicious or unskilled
users can be weeded out, providing both a better estimation of object quality
and an enhanced overall reputation-based ranking of objects.

Simulations to evaluate the effectiveness of these methods followed typical
modelling practices in physics and applied mathematics, assuming a continuum
of rating values (reflecting what may be presumed to be fine-grained shades of
opinion).  However, a near-universal feature of real user feedback and rating
systems is that they permit ratings to take only a limited range of discrete
values---most commonly the 5-star system employed by Amazon, YouTube, etc.
The influence of this constraint has never been tested on the aforementioned
algorithms, and the main purpose of the present letter is to explore how this
quantization of ratings affects the co-determination procedure and the
resulting ranking and reputation values.

Our simulations show that if the number of available rating choices is too few,
this has a strong negative impact on the algorithms' performance.
Paradoxically, in such circumstances, having a community of users more prone to
individual rating errors may actually increase the overall performance of the
system.  We compare these results with psychometric research on the measurement
of attitudes, and discuss the implications for the construction of effective
online reputation, ranking and rating systems.

\section{Algorithms}
The reputation and ranking algorithms explored in this paper all operate upon
the same basic type of data.  Suppose we have a set $\Uset$ of users who
have each rated some subset of the complete set $\Oset$ of objects.  For
notational clarity we use Latin letters ($i,j,\dots$) for user-related indices
and Greek letters ($\alpha,\beta,\dots$) for object-related indices.  The set
of users who rated a given object $\alpha$ is denoted by $\Uset_{\alpha}$,
while the set of objects rated by a user $i$ is denoted by $\Oset_i$, and
the value of the rating of object $\alpha$ by user $i$ is denoted by $r_{i\alpha}$.

We assume that each object has an intrinsic quality $Q_{\alpha}$ from which the
received ratings differ to a greater or lesser degree depending on the ability
of the user.  While in some online reputation systems there is an opportunity
for users to `rate the ratings', providing an extra measure of user reputation,
we do not rely on the availability of such information: all the algorithms
described here calculate user ability solely on the basis of the rating data.
On the basis of such measures of user ability we can then estimate object
quality using a \emph{weighted average} of the ratings,
\begin{equation}
  \label{eq:q}
  q_{\alpha} = \sum_{i\in\Uset_{\alpha}} w_{i} r_{i\alpha}\bigg/\sum_{i\in\Uset_{\alpha}} w_i \,,
\end{equation}
where the user weights $w_{i}$ are constructed by one of the following
algorithms.

(i) \emph{Arithmetic average} (\textsf{AA}).
The baseline for comparison of reputation and ranking methods is simply to
treat all user ratings equally, setting $w_{i} = \textrm{const}\ \forall\ i \in \Uset$,
or
\begin{equation}
  \label{eq:q-avg}
  q_\alpha = \frac{1}{\lvert\Uset_{\alpha}\rvert} \sum_{i\in\Uset_{\alpha}} r_{i\alpha} \,,
\end{equation}
which is of course the actual aggregation method used on most websites.

(ii) \emph{Mizzaro's algorithm} (\textsf{Mizz}).
Mizzaro~\cite{mizzaro2003} has introduced a co-determination algorithm for the
assessment of scholarly articles, with reputation scores for authors, articles
and readers that co-evolve over time according to the ratings readers give to
papers.  The algorithm can readily be applied to the more general user-object
case we consider here, with author scores omitted since their evolution is
decoupled from the evolution of article and reader scores and they are
irrelevant in the present context.  For consistency with the rest of the paper
we refer henceforth to objects and users instead of articles and readers.

The algorithm can be implemented in two versions, an \emph{incremental} one
where ratings are added one by one and an \emph{iterative} one that can be
applied to a pre-existing dataset.  We have implemented both versions (which in
any case, given the same data, produce the same result), but for ease of
comparison to the other algorithms we describe here the simpler, iterative
version.

Given a set of user weights $w_{i}$, object quality values $q_{\alpha}$ are
calculated according to Eq.~(\ref{eq:q}).  User weights are then recalculated
according to,
\begin{equation}
  \label{eq:w-Mizz}
  w_{i} = \frac {\sum_{\alpha\in\Oset_{i}} s_{\alpha} g_{i\alpha}} {\sum_{\alpha\in\Oset_{i}} s_{\alpha}}
\end{equation}
where
\begin{equation}
  \label{eq:s}
  s_{\alpha}=\sum_{j\in\Uset_{\alpha}} w_{j}
\end{equation}
is a measure of \emph{steadiness} of object quality $q_{\alpha}$, and
\begin{equation}
  \label{eq:g}
  g_{i\alpha} = 1 - \sqrt{\lvert r_{i\alpha} - q_{\alpha}\rvert / \Delta r}
\end{equation}
is a measure of disagreement between the given rating and the object score.
$\Delta r$ represents the width of the rating range, i.e.\ the difference
between the smallest and largest possible rating values, and this normalization
guarantees that the value of $g$ will fall within the range $[0;1]$.

The algorithm is initialized by setting equal weights $w_{i}=1$ for all users
$i$ and then iterating repeatedly over the equations (\ref{eq:q}, \ref{eq:w-Mizz})
until the change in the vector of quality estimates between successive
iteration steps,
\begin{equation}
  \label{eq:convergence}
  \lvert\vek{q}-\vek{q}'\rvert := \bigg[ \frac{1}{\lvert\Oset\rvert} \sum_{\alpha\in\Oset} (q_{\alpha}-q_{\alpha}')^2 \bigg]^{1/2} \,,
\end{equation}
falls below a certain threshold value\footnote{
  Note that the algorithm may fail to converge if the threshold $\Delta$ is set
  too low~\cite{dekerchove2007}.  Conversely, too large a threshold may disrupt
  the iterative process.  It may therefore take a few trials to choose an
  appropriate value.
} $\Delta$ (in our simulations, we use $\Delta=10^{-4}$).

(iii) \emph{The Yu-Zhang-Laureti-Moret algorithm} (\textsf{YZLM}).
Yu and colleagues~\cite{yu2006,*laureti2006} have introduced an algorithm that
is essentially a generalized version of maximum likelihood estimation (MLE)~\cite{fisher1925},
using a control parameter $\beta \geq 0$ to determine how divergence from the
community consensus affects user weight $w_{i}$.  Their own implementation
considers only the case where all users have rated all objects, but it is
trivial to generalize it to operate on sparse data.

Estimated object quality values $q_{\alpha}$ are again calculated according to
Eq.~\ref{eq:q}.  We then calculate the divergence between the ratings of each
user $i$ and the estimated object quality values,
\begin{equation}
  \label{eq:d}
  d_{i} = \frac{1}{\lvert\Oset_{i}\rvert} \sum_{\alpha\in\Oset_{i}} (r_{i\alpha} - q_{\alpha})^2 \,,
\end{equation}
and the updated weight of user $i$ is then given by
\begin{equation}
  \label{eq:w-YZLM}
  w_{i} = \left(d_{i} + \varepsilon\right)^{-\beta} \,,
\end{equation}
where the exponent $\beta \geq 0$ determines the strength of the penalty
applied to users with larger rating divergence $d_{i}$ (note that $\beta=0$
corresponds simply to the arithmetic average) and $0 < \varepsilon \ll 1$ is a
small positive constant so as to prevent user weights diverging (in our
simulations, we use $\varepsilon = 10^{-8}$).
Yu et al.~\cite{yu2006} noted that while $\beta=1/2$ provides better numerical
stability of the algorithm as well as translational and scale invariance,
$\beta=1$ is the optimal algorithm from the point of view of mathematical
statistics~\cite{hoel1984}.  We have used $\beta=1$ because it yields superior
performance, but choosing $\beta=1/2$ does not alter the fundamental character
of the results obtained here.

The algorithm is initialized like \textsf{Mizz}, by setting the weights
$w_{i}=1$ for all users $i$ and then iterating repeatedly over the equations
(\ref{eq:q}, \ref{eq:d}, \ref{eq:w-YZLM}) until the vector of quality estimates
$\vek{q}$ changes less than the threshold value $\Delta$.

(iv) \emph{de Kerchove and Van Dooren's algorithm} (\textsf{dKVD}).
De Kerchove and Van Dooren~\cite{dekerchove2007,*dekerchove2008,*dekerchove2009,*dekerchove2010}
have introduced an algorithm similar to \textsf{YZLM}, but where the weight
update function is given instead by
\begin{equation}
  \label{eq:w-dKVD}
  w_{i} = 1 - kd_{i} \,,
\end{equation}
where $k$ is chosen such that $w_{i} \geq 0$.  This has the advantage of
guaranteeing convergence to a unique solution independent of starting weights
(though in practice this is not a particular problem with any of the
algorithms).  In our simulations we adopt the strongest possible punishment of
noisy raters by setting $k = \left[\varepsilon + \max_{j\in\Uset} d_{j}\right]^{-1}$,
where $\varepsilon$ ensures non-zero weights if the $d_j$ are all identical.
The algorithm is initialized in a similar manner to \textsf{Mizz} and
\textsf{YZLM} by setting $w_{i}=1$ for all users $i$ and then iterating over
Eqs.~(\ref{eq:q}, \ref{eq:d}, \ref{eq:w-dKVD}) until the vector of quality
estimates $\vek{q}$ changes less than the threshold value $\Delta$.

\section{Artificial datasets}
To test the methods described above, we create artificial datasets in the
following way.  For each object $\alpha$ we randomly generate a real-valued
true quality value $Q_{\alpha}$ from the uniform distribution\footnote{
  It is possible to use non-uniform distributions, but given the limited rating
  scale this makes little practical difference.  A more pertinent question is
  whether there can actually be such a thing as a `true', objective quality
  value.  The reasonableness of this assumption will vary depending on what
  kind of objects are being considered, probably with particular reference to
  whether an object will be assessed more on the basis of taste or
  functionality.
} $U[1;R]$, where $R$ is an integer $\geq 2$.
Similarly, for each user $i$ we randomly generate a personal error
level $\sigma_{i}$ from the distribution $U[\sigma_{\min};\sigma_{\max}]$,
where $\sigma_{\min}$ and $\sigma_{\max}$ scale with the width $\Delta r = R-1$
of the rating scale.  For a given sparsity of the dataset $0<\eta\leq 1$, we
randomly select $\eta \lvert\Uset\rvert \lvert\Oset\rvert$ unique user-object
pairs\footnote{
  In the extremely rare case that an object or user ends up without any such
  links, we discount them from further consideration, e.g.\ when assessing
  algorithms' performance.
} $i\alpha$ and generate corresponding individual user estimates of object
quality according to,
\begin{equation}
  \label{random_rating}
  q_{i\alpha}=Q_{\alpha} + E_{i\alpha}
\end{equation}
where the quality estimation error $E_{i\alpha}$ is drawn from the uniform
distribution $U[-\sigma_{i};\sigma_{i}]$.  The actual ratings are derived
from these quality estimates depending on the degree of quantization desired:
for continuous-valued ratings we simply take $r_{i\alpha}=q_{i\alpha}$, while
discrete rating values are obtained by rounding to the nearest integer, that
is, $r_{i\alpha}=[q_{i\alpha}]$.  In both cases, values lying outside the
prescribed range $[1;R]$ are truncated: those smaller than $1$ are changed to
$1$ and those greater than $R$ are changed to $R$.  This follows the real-life
constraint that, no matter how much a user may adore or detest a particular
object, they still cannot rate it higher or lower than the given rating bounds.
While changing $R$ does not produce a qualitative difference in outcome for
continuous-valued ratings, the constraint of discrete integer values means that
$R$ determines the \emph{resolution} of rating precision, that is, the number
of distinct discrete rating values.  Note that since we assume $\sigma_{\min}$
and $\sigma_{\max}$ scale with $\Delta r = R-1$ this is equivalent to
increasing the resolution by taking a higher number of equally-spaced discrete
rating values within a fixed range: increasing the width of the rating scale
and taking integer values is simply easier to implement.

\begin{figure*}[t]
  \centering
  \includegraphics[width=\textwidth]{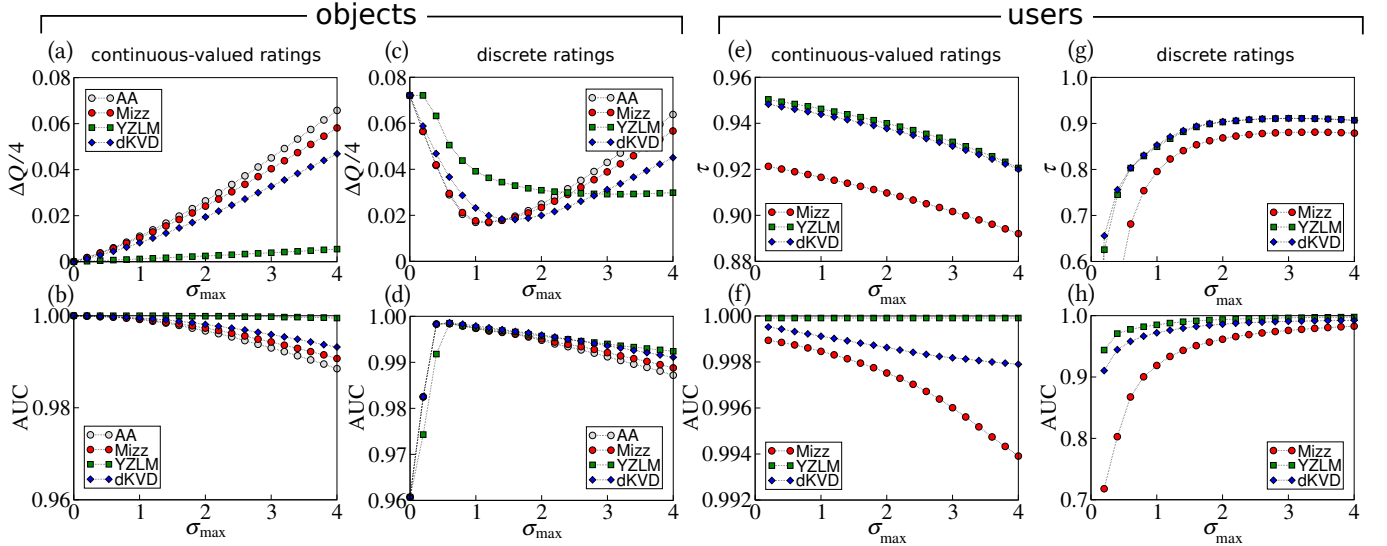}
  \caption{
    (colour online)
    Overview of co-determination algorithm performance as the upper bound
    $\sigma_{\max}$ of user error is varied, for continuous and discrete
    (integer) valued ratings in the interval $[1;5]$, with $\sigma_{\min}=0$,
    $\lvert\Uset\rvert=1000$, $\lvert\Oset\rvert=1000$, $\eta=0.1$, and results
    averaged over 100 realizations.
    (a--d)~Accuracy in estimating object quality for (a,~b)~continuous and
    (c,~d)~discrete ratings, measured by $\Delta Q$ and AUC.  For comparison to
    other figures, $\Delta Q$ is normalized with respect to the rating width
    $\Delta r=R-1=4$.
    (e--h)~Accuracy in ranking user ability for (e,~f)~continuous and
    (g,~h)~discrete ratings, measured by Kendall's $\tau$ rank correlation
    coefficient and AUC.  No results are shown for \textsf{AA}, as it does not
    rank users but considers them to all be equal.}
  \label{fig:real+int}
\end{figure*}

\section{Performance metrics}
A simple and easy test of algorithm performance is to compare the algorithm's
estimated quality values $q_{\alpha}$ and the `true' quality $Q_{\alpha}$,
and calculate the root-mean-square error~\cite{yu2006,*laureti2006},
\begin{equation}
  \label{std_error}
  \Delta Q:=\bigg[ \frac{1}{\lvert\Oset\rvert}\sum_{\alpha\in\Oset} \big(Q_{\alpha} - q_{\alpha}\big)^2 \bigg]^{1/2} \,,
\end{equation}
using the normalization $\Delta Q / (R-1)$ to compare performance on datasets
with different rating resolution $R$.  Since user weight is not expected to be
equal to the true user ability, we use Kendall's $\tau$ rank correlation coefficient~\cite{kendall1938}
to compare the true ability ranking of users according to $\sigma_{i}^{-1}$
with the estimated ranking given by $w_{i}$.  A result of $\tau=1$ indicates
that the true and estimated rankings are identical, $-1$ that they are
completely inverted, and $0$ that the rankings are entirely uncorrelated.

While well defined for artificial numerical simulations, neither of these
measures can easily be applied to real data, where objective measures of object
quality or user ranking difficult or impossible to obtain.  In the absence of
reliable per-item or per-user measures of accuracy, an effective approach is to
specify a group of `relevant' objects or users and inspect their position in
the ranking~\cite{herlocker2004}.  To do this we employ the receiver operating
characteristic (ROC) curve~\cite{swets1963,*hanley1982,*fawcett2006},
constructed by plotting for each place in the ranking a point in $[0,1]^{2}$
whose $x, y$ values correspond respectively to the proportion of irrelevant and
relevant objects recovered so far.  The ranking accuracy can then be estimated
by the area under the curve (AUC), which equals $1$ when every relevant
object/user is ranked higher than every irrelevant object/user, $0.5$ when the
distribution of relevant objects/users is random, and $0$ when every irrelevant
object/user is ranked higher than every relevant object/user.  In the
simulations presented here, we denote as `relevant' the $5\%$ of objects/users
with respectively the highest true quality values $Q_{\alpha}$ or lowest error
$\sigma_{i}$.

\section{Results}
For the results presented here we generated artificial datasets of $1000$ users
and $1000$ objects, with sparsity $\eta=0.1$.  For each simulation we used the
same datasets to test each reputation algorithm.  Our first simulations keep
a constant rating resolution $R=5$ and a constant lower bound $\sigma_{\min}=0$
for the distribution of user's personal error levels, while the upper bound
$\sigma_{\max}$ was varied in the range $[0;4]$.  This range was chosen so
that, at its most extreme, the least skilled users (i.e.\ those with $\sigma_{i}
\approx \sigma_{\max}$) could potentially rate a `perfect' 5-star object with
the lowest rating value 1, and vice versa.

\begin{figure}
\centering
\includegraphics[width=0.49\textwidth]{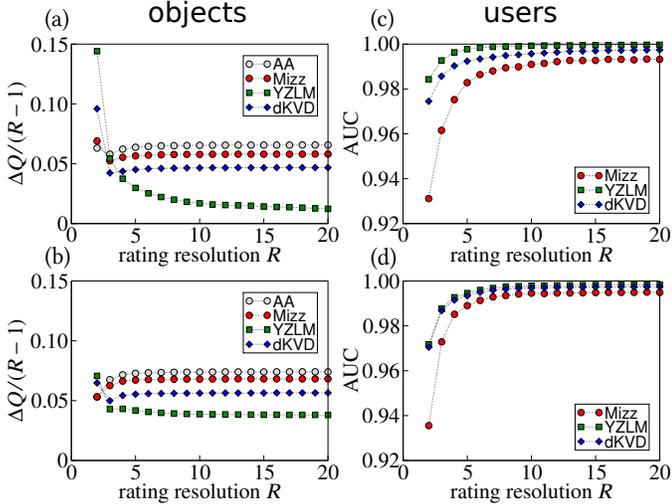}
\caption{
  (colour online)
  The dependency of algorithm performance on the discrete rating resolution $R$,
  measured by (a,~b)~$\Delta Q$ for objects and (c,~d)~AUC for users.
  The upper error bound $\sigma_{\max}=R-1$ covers the full rating range, while
  lower error bounds are (a,~c)~$\sigma_{\min}=0$, (b,~d)~$\sigma_{\min}=\sigma_{\max}/8$.
  Other parameter values are as in Fig.~\ref{fig:real+int}.
}
\label{fig:scale}
\end{figure}

Figure~\ref{fig:real+int}a,b,e,f presents the performance of the algorithms
when we use continuous-valued ratings, i.e.\ when $r_{i\alpha} = q_{i\alpha}$
exactly, and vary the upper error bound $\sigma_{\max}$.  We observe
immediately that \textsf{YZLM} is by far the least sensitive to the increasing
error level, maintaining the lowest object quality error $\Delta Q$ and the best
user ranking (Kendall's $\tau$), and the highest AUC ($\approx 1$) for both
objects and users.  This is because of all the methods \textsf{YZLM} places the
harshest sanction against `noisy' raters who diverge from the aggregate
estimated quality, a feature that can be observed in the ranking of users,
where we observe near-identical values of $\tau$ for both YZLM and dKVD (their
weights and hence ranking stem from the same measure $d_{i}$ of user rating
divergence) but consistently higher AUC for YZLM (the very best users are more
consistently pushed to the top of the ranking due to the harsher sanction).
The superiority of \textsf{YZLM} is maintained across different sizes of
dataset and different data sparsity values, and is found to be dependent
primarily on $\sigma_{\min}$: if this lower error bound is increased, results
from all four algorithms become similar as, in the absence of objectively good
raters, there is much less advantage to be had in discriminating between better
and worse\footnote{
  The degree of superiority shown by \textsf{YZLM} actually depends both on the
  value of $\sigma_{\min}$ and the difference $\sigma_{\max}-\sigma_{\min}$.
  We do not provide a detailed illustration of this for reasons of space, but
  the effect can be observed in the differences in asymptotic values of AUC
  between the upper and lower panels of Fig.~\ref{fig:scale}.
}.

To assess the difference between continuous- and discrete-valued ratings, we
took the same sets of artificial data and repeated the analysis with ratings
now constrained to integer values (1--5).  As shown in Fig.~\ref{fig:real+int}c,d,g,h,
this quantization has a substantial negative effect on performance, with
$\sigma_{\max}=0$ in particular being disastrous for all reputation algorithms.
As $\sigma_{\max}$ increases, $\Delta Q$, $\tau$ and object and user AUC improve---and
then, in some cases, $\Delta Q$ and object AUC worsen again.  We also notice
that the relative performance of the methods with respect to $\Delta Q$ and
object AUC is inverted for $\sigma_{\max}<2$, with \textsf{YZLM} the
worst-performing of the algorithms, regaining its superiority only when the
upper error bound is large.

The apparent paradox of better performance resulting from increasing error can
be explained as follows. Imagine an object with `true' quality $3.4$ being
assessed by two distinct groups of users, the first whose quality assessment is
always error-free ($\sigma_i=0$), the second whose error levels are set at
$\sigma_i=0.5$ (i.e.\ the average error level of a user from a group with error
levels $\sigma_i$ drawn from $U[0;1]$).  Users from the first group will of
course make correct quality judgements $q_{i\alpha}=Q_{\alpha}$, but the
discrete rating system forces them to adopt the nearest integer value of $3$.
The resulting average (also $3$) will thus differ from the true quality by
$0.4$.  By contrast the `noisy' users' quality estimates will be distributed
uniformly in the range $Q_{\alpha}\pm 0.5$ and so on average 60\% of them will
give a discrete rating of 3 and 40\%\ will give 4, leaving an average of
3.4---that is, on average a perfect match to the original quality value.
Effectively, the constraint of discrete ratings produces a systematic
quantization error, which `noisy' users can offset in the same way that dither
can reduce quantization error in signal processing~\cite{roberts1962,*gray1993,*carbone1994}.

A slightly more subtle argument is needed to explain the bad performance of
\textsf{YZLM} when faced with any but the largest levels of error.  Here we
note that, while the \emph{aggregate} error of low-$\sigma_i$ agents may be
greater, their \emph{individual} error will still on average be less.
\textsf{YZLM}, with its strong bias towards users with low observed error
rates, will thus favour these users, suppressing noisy agents and consequently
harming aggregate performance.  This is confirmed by Fig.~\ref{fig:real+int}g,h,
where we observe that while overall accuracy of user ranking (Kendall's $\tau$)
suffers with low $\sigma_{\max}$, the high AUC values indicate that the
lowest-$\sigma_{i}$ users are still being pushed towards the top of the
ranking.  As $\sigma_{\max}$ increases, aggregate error of the wider population
grows and \textsf{YZLM}'s suppression of high individual error rates acts to
suppress this, sustaining its performance while the other algorithms suffer.

To better understand the effects of changing the rating resolution, we
performed simulations where user error was fixed in proportion to the width of
the rating scale, and varied the value of $R$ while taking discrete ratings.
Fig.~\ref{fig:scale} shows the results for two sets of simulations, the first
with $\sigma_{\min}=0$, the second with $\sigma_{\min}=\sigma_{\max}/8$, using
$\Delta Q/(R-1)$ and AUC to measure performance in assessing objects (a,~b) and
users (c,~d) respectively.  In both cases $\sigma_{\max} = R-1$, so that the
maximum possible user error covers the full range of the rating scale.

As we increase the rating resolution $R$, we observe a gradual approach to
asymptotic values of object ($\Delta Q$) and user (AUC) performance
comparable to those obtained with continuous-valued ratings.  Similar to
Fig.~\ref{fig:real+int}c, there is a marked difference between \textsf{YZLM}
and the other algorithms with respect to $\Delta Q$.  Whereas \textsf{AA},
\textsf{Mizz} and \textsf{dKVD} have only a limited response to increasing
resolution, \textsf{YZLM} is able to reap a significant benefit, with its
performance sustaining continuous improvement even as $R$ approaches $20$.  The
reason is made clear by Fig.~\ref{fig:scale}c, where we observe that unlike the
other algorithms, increasing $R$ permits \textsf{YZLM} to push the lowest-$\sigma_{i}$
users consistently to the very top of the ranking ($\textrm{AUC}\to 1$).

\textsf{YZLM}'s dependency on low-$\sigma_{i}$ raters is further emphasized by
Fig.~\ref{fig:scale}b, where the performance of \textsf{AA}, \textsf{Mizz} and
\textsf{dKVD} are little affected by the higher value of $\sigma_{\min}$ but
where now \textsf{YZLM} performs better for binary ratings (again, the
`increased noise=better performance' paradox) while no longer sustaining any
significant improvements in $\Delta Q$ for $R>3$.  When observing AUC for user
ranking (Fig.~\ref{fig:scale}d), we observe that now \textsf{YZLM} too is
unable to consistently push the lowest-$\sigma_{i}$ users to the top, with asymptotic
AUC values now near-identical to \textsf{dKVD}.

\section{Discussion}
Psychometric research has put considerable effort into understanding the
effectiveness and reliability of different rating scales, particularly with
respect to the scale resolution~\cite{green1970,*cox1980,*alwin1992,jacoby1971,*benson1971,lehman1972,preston2000,*svensson2000}.
Factors to take into account include both the information-carrying capacity of
the scale and the information-\emph{processing} capacity of respondents~\cite{hulbert1975},
as well as psychological influences such as the descriptive labels associated
with responses~\cite{alwin1991}.

The relevance of these factors depends on exactly what kind of information one
wants to extract from the scale.  If the aim is to aggregate or average over
respondents, three or even two discrete response options may suffice~\cite{jacoby1971,*benson1971}.
Conversely, if the focus is on individual difference, finer-grained scales
become necessary~\cite{lehman1972}.

Co-determination algorithms are \emph{prima facie} aggregation mechanisms, but
they also employ measures of individual difference to improve the aggregation
process~\cite{mizzaro2003,yu2006,*laureti2006,dekerchove2007,*dekerchove2008,*dekerchove2009,*dekerchove2010}.
The effect of rating resolution on their performance will therefore depend on
several factors, including the degree to which there are meaningful and
reliable differences in user rating ability, whether the scale is fine-grained
enough to accurately reflect those differences, and the algorithm's ability to
measure and exploit this information if it exists.

In this letter we have investigated the influence of low rating resolution on
the performance of several co-determination reputation and ranking algorithms.
While the presence of an non-zero optimal noise level (Fig.~\ref{fig:real+int}c,d)
may be seen as a mere mathematical curiosity---in effect an example of
quantization error being reduced by the application of dither~\cite{roberts1962,*gray1993,*carbone1994}---the
worsened performance of these methods is an important finding.  Psychometric
studies have in general suggested that there is little benefit to be had from
using more than 7 discrete rating categories~\cite{green1970,*cox1980,*alwin1992}.
Our results suggest that in fact this may prevent the maximum exploitation of
rating data, precluding the fine-grained observation of individual difference
necessary to improve the aggregation process.  A comparison can be drawn to
models of opinion dynamics inspired by the Potts model, where if the number of
spin/opinion values is too few, opinions become homogenized across the
population, while as $q\to\infty$, diverse regions of different opinion can be
preserved~\cite{wu1982,*sire1995,*axelrod1997,*castellano2009}.

We have also shown that, where the rating resolution is high enough,
co-determination algorithms---particularly \textsf{YZLM}---are able to achieve
significantly better results than a mere arithmetic average.  Given that
psychometric studies have not shown any major \emph{disadvantages} of using
higher-resolution scales~\cite{preston2000,*svensson2000}, it may thus be
preferable for modern rating and reputation systems to employ continuous-valued
scales such as the graphic rating scale or the visual analogue
scale~\cite{hayes1921,*freyd1923a,*freyd1923b,*aitken1969,*ahearn1997}. In an
online world such scales can be implemented easily through the use of
percentage scores or slider bars~\cite{marsh-richard2009,*ladd2009}.  Empirical
studies employing these and other rating methods should be able to determine if
and when respondents are in practice able to achieve the required precision of
judgement, and so help to identify the situations where a sophisticated method
may yield superior performance.

\acknowledgements

  We thank Yi-Cheng Zhang, Yi-Kuo Yu, Hassan Masum, Tao Zhou and Luo-luo Jiang
  for many inspiring conversations, and two anonymous referees for their
  valuable feedback.  Special thanks to Judith Simon and Ethan Munson for
  suggesting connections to the psychometrics and signal processing literature.
  This work is part of the Liquid Publications Project (\url{http://project.liquidpub.org/})
  under EU FET-Open grant no.\ 213360.

\bibliographystyle{eplbibM}
\bibliography{medo-wakeling_discrete-continuous}

\end{document}